\documentclass[review,5pt,twocolumn,preprint]{elsarticle}
\usepackage{lineno,hyperref}
\modulolinenumbers[5]
\usepackage{amsmath}

\begin{document} 
\twocolumn[{\begin{frontmatter}
%\begin{frontmatter} 
    
\title{Thermoelectric Transport in Graphene/{\it h}-BN/Graphene Heterostructures: A Computational Study}

%\author{Ransell D'Souza, Sugata Mukherjee}
%\address{S.N. Bose National Centre for Basic Sciences, Block JD, Sector III, Salt Lake, Kolkata 700098, India}
\author[mymainaddress]{Ransell D'Souza}
\ead{ransell.d@gmail.com}

\author[mymainaddress]{Sugata Mukherjee\corref{mycorrespondingauthor}}
\cortext[mycorrespondingauthor]{Corresponding author}
\ead{sugata@bose.res.in; sugatamukh@gmail.com}

\address[mymainaddress]{S.N. Bose National Centre for Basic Sciences, Block JD, Sector III, Salt Lake, Kolkata 700098, India}

\begin{abstract}
We present first principles study of thermoelectric transport properties of sandwiched heterostructure of Graphene (G)/hexagonal Boron Nitride (BN)/G, 
based on Boltzmann transport theory for band electrons using the bandstructure calculated from the Density Functional Theory (DFT) based plane-wave method.  
Calculations were carried out for three, four and five BN layers sandwiched between Graphene layers with three different arrangements to obtain the Seebeck
coefficient and Power factor in $T\sim 25-400$K range.
Moreover, using Molecular Dynamics (MD) simulations with very large simulation cell we obtained the thermal conductance ($K$) of these heterostructures 
and obtained finally the Figure-of-Merit ($ZT$). These results are in agreement with recently reported experimental measurements.

\end{abstract}


\end{frontmatter}
}]
%*Corresponding author (e-mail: sugata@bose.res.in; sugatamukh@gmail.com)

%\maketitle

\section{Introduction}

The study of thermoelectric transport in nanomaterials has been a topic of intensive research in recent years \cite{dresselhaus12, majumdar04, cahill14}.
Nanomaterials in the form of compound semiconductors, semiconductor multilayers and superstructures offer possibilities to exhibit enhanced
thermoelectric Figure-of-Merit ($ZT$) caused by the simultaneous decrease in  thermal conductivity ($\kappa$) and increase in electrical conductivity ($\sigma$)
of the material. This has lead an intensive effort to search for novel materials useful for energy research.
  
Graphene (G) and Hexagonal boron nitride ({\it h}-BN) atomically layered architectures are potential candidates for device applications. 2-D graphene 
transistors based on lateral heterobarriers have been proposed and investigated from the theoretical point of view \cite{fiori11, fiori12}, which were motivated by 
the first experimental success in realizing G-BN lateral heterostructures \cite{ci10}. Based on simulation studies, vertical heterobarrier graphene transistors have 
also been proposed \cite{sciambi11}, \cite{mehr12}. Britnell et. al. \cite{britnell12} have reported a prototype field-effect tunneling transistor with atomically thin 
boron nitride acting as a vertical transport barrier. In addition to electron transport, heat dissipation in these graphene based heterojunction devices is found to be 
dominated by vertical heat transfer \cite{jo11, pop12}. Recently Chen et al \cite{chung14} have reported thermoelectric transport measurements across G/{\it h}-BN/G 
heterostructure with multiple {\it h}-BN layers. Based on the observed thermoelectric voltage and temperature gradient, they have obtained a Seebeck coefficient ($S$) 
of $\textendash$99.3 $\mu$V/K and 
Thermoelectric Power Factor ($S^2\sigma)$ of 1.51$\times 10^{-15}$ W/K$^2$ for the heterostructure device, respectively. From the thermal transport measurements 
of Jo et al \cite{jo13}, thermal conductivity of multilayers of {\it h}-BN was obtained and finally the Figure-of-merit ($ZT$) of G/{\it h}-BN/G heterostructure 
was estimated \cite{chung14} to be 1.05$\times 10^{-6}$.

Quantum transport in trilayers G/{\it h}-BN/G and {\it h}-BN/G/{\it h}-BN have recently been investigated theoretically by Zhong et al \cite{pandey12}, predicting a
metal like conduction in low-field regime. Thermoelectric transport in G/{\it h}-BN nanoribbons have been studied using non-equilibrium Green's function method 
\cite{rubio2012} for different thickness of {\it h}-BN and Graphene domains. Thermal transport was studied by Kinaci et al \cite{kinaci} in G/{\it h}-BN nanoribbons
using equilibrium molecular dynamics simulation. However, a complete study of thermoelectric transport in G/{\it h}-BN/G heterostructures with multilayer {\it h}-BN 
and comparison to experimental data \cite{chung14} is apparently not available.   

In this paper we report a computational study of the thermoelectric properties of sandwiched heterostructures of Graphene and {\it h}-BN. We have used density functional 
theory (DFT) based electronic structure method and Boltzmann transport theory for the band electrons to calculate the electrical conductivity ($\sigma$) and 
Seebeck coefficient ($S$). A large-scale equilibrium molecular dynamics (MD) simulation using Green-Kubo formalism \cite{green, kubo} at constant temperatures 
was used  to compute thermal conductivity ($\kappa$) of these heterostructures at various temperatures. Our calculations allow us study of electrical and thermal
transport in the directions parallel and normal to the plane of G/{\it h}-BN/G and thus permits a direct comparison of our simulation results to the experimental data. 
Our calculations show that for certain configuration of the heterostructured nanomaterials the Power factor and Figure-of-merit ($ZT$) are close to recent measurements 
\cite{chung14}. Moreover, our calculated $\kappa$ for the multilayers and bulk {\it h}-BN shows a qualitative agreement with recent experimental results of Jo et al 
\cite{jo13}. Calculated $\kappa$ along orthogonal directions in planar G/{\it h}-BN striped heterostructures also quantitatively agree with previous calculations 
\cite{kinaci}. 

\section{Method of Calculation}

\begin{figure}[!htbp]
\includegraphics[scale=0.2]{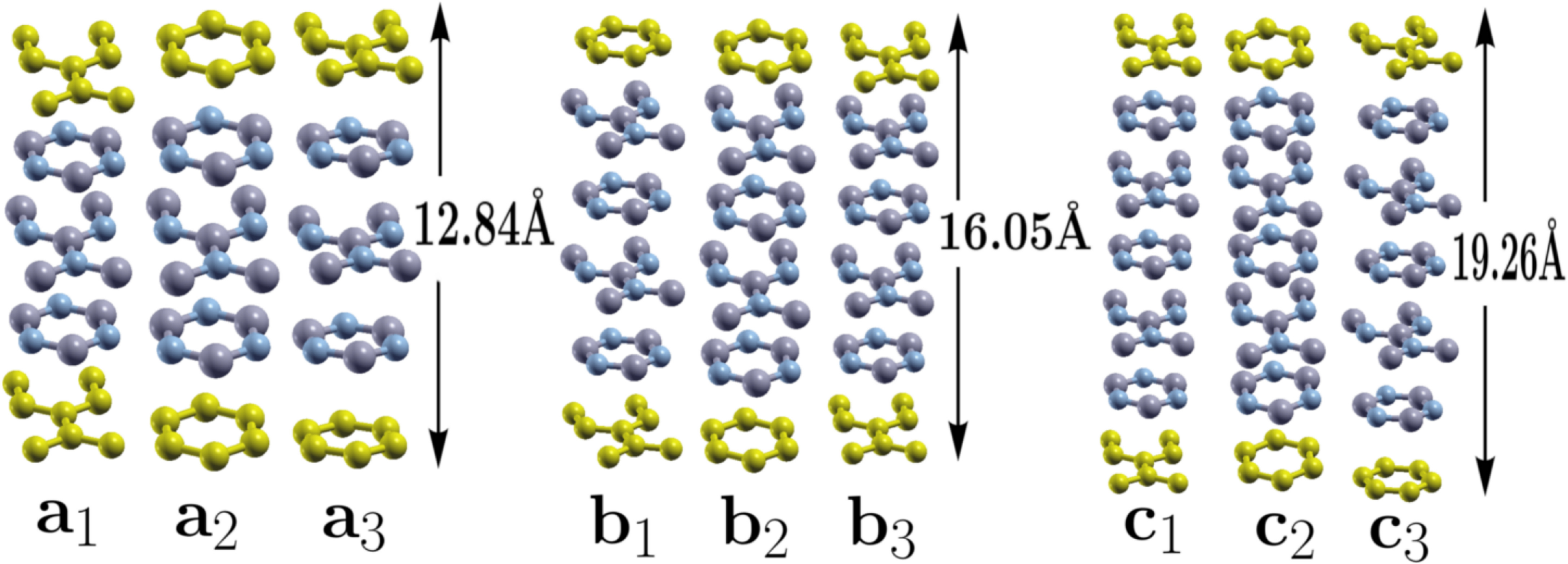}
\centering
\caption{\label{difftype} Supercells of G/{\it h}-BN/G heterostructures with three (left), four (middle) and five (right) {\it h}-BN layers showing three different types 
of arrangement of Graphene and {\it h}-BN layers. The numbers indicate thickness of the heterostructures in {\AA}.}
\end{figure}

All the electronic structure calculations were carried out using Density Functional Theory (DFT) based plane-wave method, as implemented in the Quantum Espresso code 
{\cite{giannozzi09}}, using an orthorhombic unit-cell. The generalized gradient approximation (GGA) \cite{pbe96} was used for the exchange-correlation potential and 
the ultrasoft 
pseudopotential \cite{vanderbilt90} was used to describe the core electrons. Self-consistent calculations were performed using a converged Monkhorst–Pack k-point grid
\cite{mp76} of $48\times48\times2$ with a plane wave basis with kinetic energy cutoff of 40Ry and charge density energy cutoff of 160Ry, respectively. 
The periodically repeated unit cells are separated by a vacuum spacing of 22{\AA} along the $z$-direction. This is reasonable since, the widths are typically $10^4$ 
times larger than the height of the sample \cite{jo13}. 
We have considered Van der Waals interaction \cite{grimme2, grimme3} between the layers. 
For one {\it h}-BN layer sandwiched between two graphene layers, we have minimized the total energy and pressure to get a lattice constant of 2.48{\AA} 
and interlayer distance of 3.21{\AA}. {\it h}-BN layers were added at a distance of 3.21{\AA} above the previous layer. It should be noted that addition of layers does not 
change the pressure of the unit cell. As it has been shown that the AB stacking is the most stable \cite{sm11} the {\it h}-BN layers were fixed to the AB stacking 
while graphene sheets were changed as shown in fig(\ref{difftype}).

\subsection{Electrical Conductivity and Thermopower}
To calculate the transport properties, we have used the semi-classical Boltzmann transport theory applied to band electrons as implemented by the Boltztrap 
code \cite{boltztrap1}. 
The transport parameters are obtained from the group velocity $v_{\alpha}(i,{\bf k})$ of the band electrons, referring to the i\textsuperscript{th} energy band and 
the $\alpha$ \textsuperscript{th} component of the wavevector {\bf k}, from the band dispersion $\epsilon(i,{\bf k})$ as,
\begin{eqnarray}\label{t1}
v_{\alpha}(i,\textbf{k})=\frac{1}{\hbar}\frac{\partial \varepsilon_{i,k}}{\partial k_{\alpha}}
\end{eqnarray}

The electrical conductivity tensor is then written as,
\footnotesize{\begin{eqnarray}
{\sigma_{\alpha\beta}(T,\mu) \over \tau} &=& {1 \over V} \int e^2 v_{\alpha}(i,{\bf k})\, v_{\beta}(i,{\bf k}) [{-\partial f_\mu(T,\epsilon) \over \partial \epsilon}] d\epsilon
\end{eqnarray}}

Here, $f_{\mu}$ is the Fermi-Dirac distribution function, $V$ is the sample volume and $\tau$ is the electron relaxation time, which depends on the electron-electron 
interaction and $e$ is the electronic charge.
Though one would expect that $\tau$ would depend on both the band index and $\textbf{k}$, detailed studies \cite{wwschulz, pballen2} have shown, 
to a good approximation $\tau$ could be independent of direction. Above relation was used to calculate temperature dependent resistivity of two-dimensional CBN nanomaterials
recently \cite{rdsm15}.
%Using the DOS analogy, energy projected conductivity tensors can be defined using the conductivity tensors \ref{eq2}.
%\begin{eqnarray}
%\sigma_{\alpha \beta}(\varepsilon)= \frac{1}{N}\sum_{i,\textbf{k}}\sigma_{\alpha \beta}(i,\textbf{k})\frac{\delta(\varepsilon - \varepsilon_{i,\textbf{k}})}{d\varepsilon}
%\end{eqnarray}
%where N is the number of ${\textbf{k}}$ points sampled.
%To incorporate temperature ($T$) and chemical potential ($\mu$) into $\sigma$ we use the Fermi-Dirac distribution $f_{\mu} = \frac{1}{1+ e^{\beta (E-\mu)}}$ ,
%\begin{eqnarray}\label{cond}
%\sigma_{\alpha \beta}(T;\mu)= \frac{1}{\Omega}\int \sigma_{\alpha \beta}(\varepsilon)\Bigg(\frac{-\partial f_{\mu} (T;\varepsilon)}{\partial \varepsilon}\Bigg)d\varepsilon
%\end{eqnarray}

The Seebeck coefficient tensor is expressed as,
\begin{eqnarray}\label{S}
%S_{ij}=(\sigma^{-1}_{\alpha i})v_{\alpha j}
S_{\alpha \beta}(T,\mu) = \frac{1}{eT}\frac{\int v_{\alpha}(i,{\bf k}) v_{\beta}(i,{\bf k})(\epsilon - \mu) [{-\partial f_\mu(T,\epsilon) \over \partial \epsilon}]d\epsilon}
{\int v_{\alpha}(i,{\bf k}) v_{\beta}(i,{\bf k}) [{-\partial f_\mu(T,\epsilon) \over \partial \epsilon}]d\epsilon}
\end{eqnarray}

Calculated bandstructure of G/{\it h}-BN/G heterostructure \cite{supp} indicates that this material is a narrow band gap semiconductor with a band gap $\sim 0.05$ eV. 
%It is know that for metals and semimetals, transport occurs only near the Fermi level, justifying the Sommerfeld expansion of eq. \ref{S} to get,
For the electrical transport around the energy gap, one can obtain a simpler form of the Seebeck coefficient in Eq. \ref{S}, by using the Sommerfeld expansion, to obtain,
\begin{eqnarray}\label{Smott}
S = -\frac{\pi^2 k_B^2T}{3e} {d \over dE}\, [ln\, \sigma(E)].
\end{eqnarray}

Above relation is known as the Mott formula \cite{mott}, where $k_B$ is the Boltzmann constant.
The Power factor is defined as $S^2\sigma$ and the Figure of merit is defined is defined as
\begin{eqnarray}\label{ZT}
ZT = \frac{S^2\sigma T}{\kappa},
\end{eqnarray}
where $\kappa$ is the temperature dependent thermal conductivity.

\subsection{Thermal Conductance}

In order to obtain the instantaneous heat current ($\textbf{J}$) as a function of time, one can employ equilibrium molecular dynamics simulations. Moreover, using 
this heat current, thermal conductivity $\kappa$ can be evaluated by using the Green-Kubo method \cite{green,kubo} or the Einstein relation \cite{zwanzig}.
Detailed calculations using the latter method have been reported \cite{kinaci2,kinaci3,kinaci4,kinaci5,kinaci6}. 
Here we adopt the Green-Kubo method as implemented in the code LAMMPS \cite{lammps}. 
Thermal conductivity is defined as the coefficient that links the macroscopic heat current to the temperature gradient, $\textbf{J} = -\kappa \bigtriangledown T$. The formula for $\kappa$ by Green Kubo is given by,
\begin{eqnarray}\label{kappa-gk}
\kappa = \frac{1}{3Vk_BT^2}\int\limits_0^\infty \langle \textbf{J}(0) \textbf{J}(t)\rangle \textrm{d}t,
\end{eqnarray}
where $V$ and $T$ are the volume and temperature. The factor 3 accounts for averaging over the 3 dimensions and the angular brackets refers to the average over time.
The macroscopic heat current is given by,
\begin{eqnarray}\label{j}
\textbf{J}(t) = \sum_{\substack{i}} \textbf{v}_ie_i + \frac{1}{2}\sum_{\substack{i<j}}\textbf{r}_{ij}\big(\textbf{F}_{ij}\cdot(\textbf{v}_i + \textbf{v}_j)\big),
\end{eqnarray}
where $\textbf{v}_i$ and $e_i$ are the velocity and site energy of particle $i$. $\textbf{F}_{ij}$ is force on the atom $i$ due to its neighbor $j$ from the potential.

Molecular dynamics (MD) simulations were performed first in microcanonical ensemble ($NVE$) and then the Nos\'e-Hoover ($NVT$) ensemble. The constant temperature $NVT$ 
ensemble requires an additional frictional term \cite{tomanek2000} to be introduced in Eq \ref{j}. To ensure energy conservation of the system, MD simulations were 
carried out with a time step of 1fs. Five different initial uniform seed velocities were used for the simulations. The value of $\kappa$ was obtained by averaging over these 
runs using the standard deviation as the error bar. Preliminary calculations with  different number of atoms showed that the standard deviation (error bar) was small 
when the number of atoms in the simulation cell were around 40000. Therefore, in all MD runs we used simulation cells containing around 40000 atoms. The results of MD 
simulations are shown in the next section.

We used the Tersoff potential based force field \cite{tersoff}, as obtained for {\it h}-BN
by Sevik et al. \cite{kinaci3}, and for graphene by Lindsay et al. \cite{lindsay}. The Tersoff-parameters for the B-C and N-C bonds were taken from Kinaci et al
\cite{kinaci}. To examine this Tersoff-type potential implemented in LAMMPS, we
reproduced the results of Kinaci \cite{kinaci} for the thermal conduction in two-dimensional G/{\it h}-BN stripes by calculating the perpendicular and parallel components of
the thermal conductivity in the plane of G/{\it h}-BN both for the zigzag and armchair interfaces between Graphene and {\it h}-BN domains.
Our calculated value of $\kappa/\kappa_0$, $\kappa_0$ being Thermal Conductance of pristine Graphene, compares very well with recent results obtained from non-equilibrium
Greens function method \cite{rubio2012}.

\section{Results and Discussions}
\subsection{Electrical Conductance and Thermopower}

\begin{figure}[!htbp]
\includegraphics[scale=0.3]{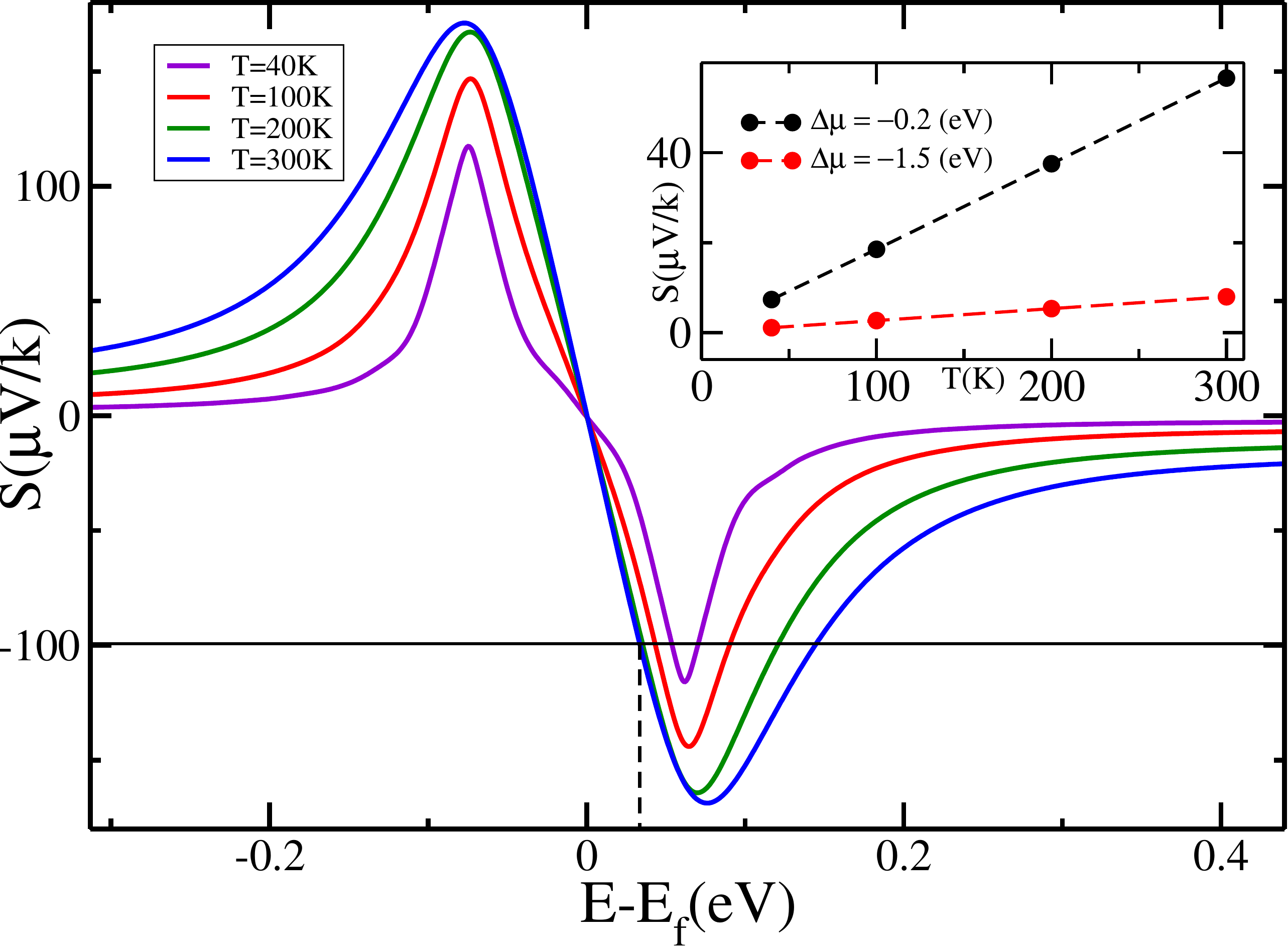}
\centering
\caption{\label{Smott-fig} Calculated Seebeck coefficient of G/{\it h}-BN/G plotted against energy using the Mott's formula using equation \ref{Smott} at various temperatures.
The black horizontal line refers to the experimental value at $T$=300K \cite{chung14}. The inset shows the Seebeck coefficient plotted against Temperature at a constant chemical potential. }
\end{figure}

\begin{figure}[!htbp]
\includegraphics[scale=0.23]{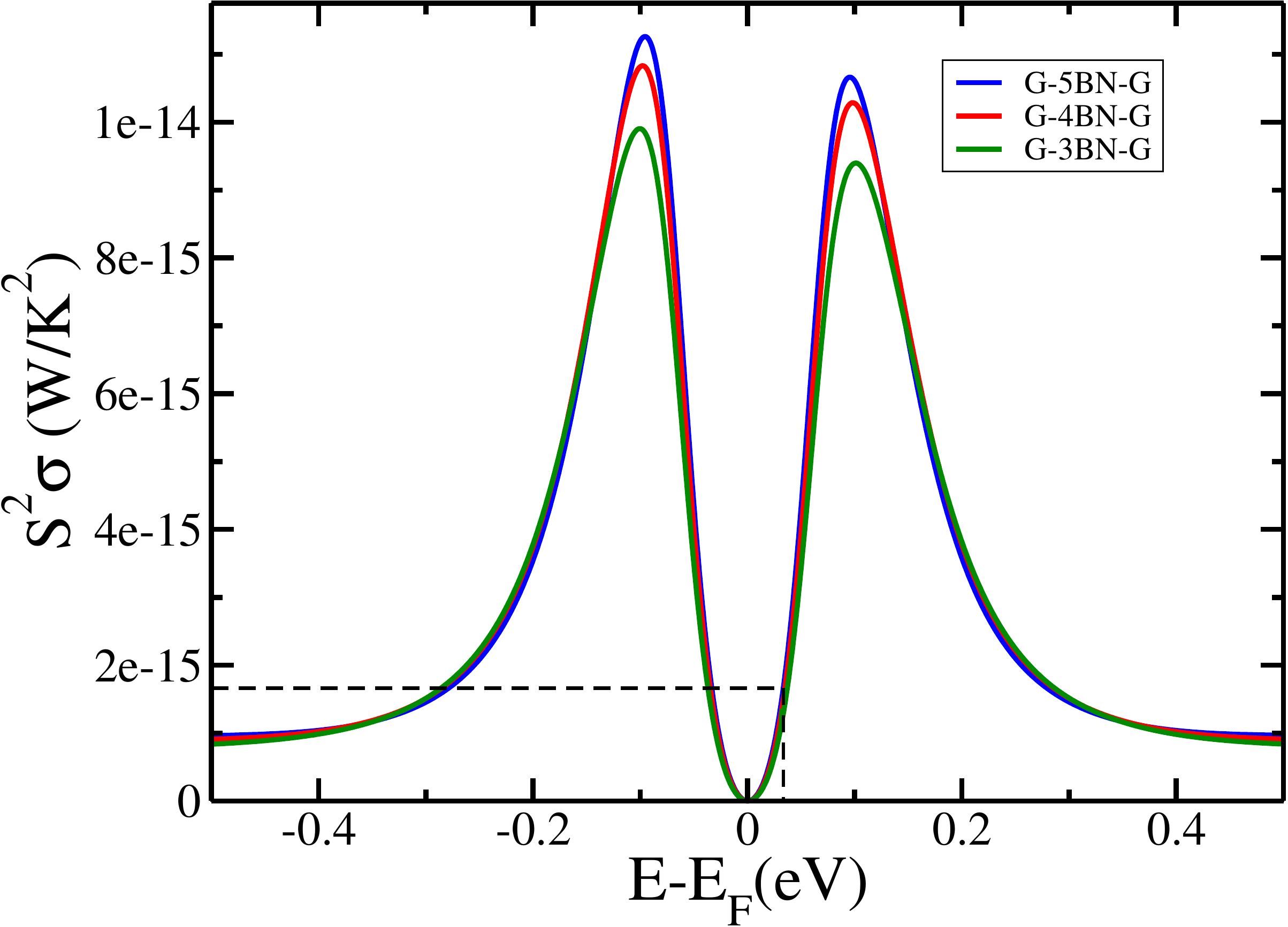}
\centering
\caption{\label{pf} Calculated Powerfactor for different layers in G/{\it h}-BN/G. The black horizontal line refers to the powerfactor which corresponds to the chemical 
potential that yields the experimental Seebeck coefficient by Chen et al \cite{chung14}.}
\end{figure}

The relaxation time for G/{\it h}-BN/G heterostructures are not known but are typically in the order of $10^{-14}$sec. Hence the numerical value used here is $1 \times 10^{-14}$sec. Calculations were performed for all topologies as shown in Fig \ref{difftype} but the results are almost identical except for a small change in Fermi energy ($E_F$). We therefore report Seebeck coefficient, Power Factor and Figure of Merit for only a specific arrangement as shown in Fig.\ref{Smott-fig} a3,b3,c3.
Fig. \ref{Smott-fig} and \ref{pf} refers to the Seebeck coefficient and power factor of G/{\it h}-BN/G heterostructure having 5 BN layers as shown in Fig.\ref{Smott-fig} c3.

Though the band gap of G/{\it h}-BN/G heterostructure is formed due to {\it h}-BN, the Fermi level does not shift since the number of boron atoms is equal to the number of nitrogen atoms. Further, since boron is an acceptor whereas nitrogen is a donor, the total number of charge carriers $n$ remains the same as that in graphene and hence the conductance at low temperatures, which is essentially proportional to $\sqrt{n}$  is essentially that of graphene \cite{kim2}. We therefore expect that the form of $S$, which depends only on conductivity, to have a similar form as that of graphene. In Fig. \ref{Smott-fig} we see that, as expected, the form of $S$ is that of graphene. However, at very low temperatures the Seebeck coefficient has a flat region around the Fermi energy which is due to the band gap.

In the 40K - 300K temperature ($T$) range it is seen that the conductivity decreases as $T$ increases. Therefore $S$ increases as $T$ increases as seen in the inset of Fig \ref{Smott-fig}, where we plot $S$ as a function of $T$ at constant chemical potential. The linear dependence of $S$ on $T$ suggests that the mechanism for thermoelectric generation is diffusive thermopower \cite{mahan02}. 
$S > 0$  for chemical potentials lower than the Fermi energy, $S = 0$ at Fermi energy and $S < 0$  for values greater the Fermi energy.
The sign of $S$ indicates the sign of the majority charge carriers. This is also observed experimentally when the gate voltage crosses the charge neutrality point (CNP). 
$S=0$  at the CNP. We thus see a direct correspondence between chemical potential and gate voltage. Thus, the gate voltage can tune the chemical potential. As seen in 
Fig \ref{Smott-fig}, the chemical potential changes sign at the Fermi level. The effect of chemical potential on $S$ can thus be demonstrated by tuning the chemical 
potential (Fermi energy $E_F$).

We would like to mention that we have calculated the the components of $\sigma$ along the cartesian axes, with $z$-axis being normal to the plane of the G/{\it h}-BN/G
heterostructure, shown in \cite{supp}. Then using the Mott's formula \ref{Smott} we obtained the Seebeck coefficients along the principal directions \cite{supp}. We have
obtained a finite $S_z$ near the Fermi energy which contributes to the total Seebeck coefficient which could be due to periodic boundary condition. We feel that the electrical conduction along the $z$-axis should
include contributions also from the other two principal directions, as planar Graphene is used as the contact on both sides of multi-layer {\it h}-BN in the experiment
\cite{chung14}.

Experimentally the Seebeck coefficient of G/{\it h}-BN/G was measured by applying a temperature gradient between the top and bottom Graphene layers using Raman
spectroscopy \cite{chung14}. For a temperature gradient $\Delta T=39$ K at a constant thermoelectric voltage $\Delta V=4$mV, they obtained $S = -99.3 \mu$V/K. 
This method was employed by Chen et al \cite{chen14} to measure the Seebeck coefficient of G/{\it h}-BN. To compare our calculations with that of Chen et al 
\cite{chung14} we fixed the chemical potential corresponding to the experimentally measured $S$ as shown in Fig. \ref{Smott-fig}.
The power factor for the three different arrangements of G/{\it h}-BN/G heterostructures are shown in Fig. \ref{pf}. The chemical potential which corresponds to that 
of the gate voltage of Chen et al \cite{chung14} has been shown by the black dotted line. 

\subsection{Thermal Conductance}
\subsubsection{G/{\it h}-BN Planar striped heterostructure}

\begin{figure}[!htbp]
\includegraphics[scale=0.3]{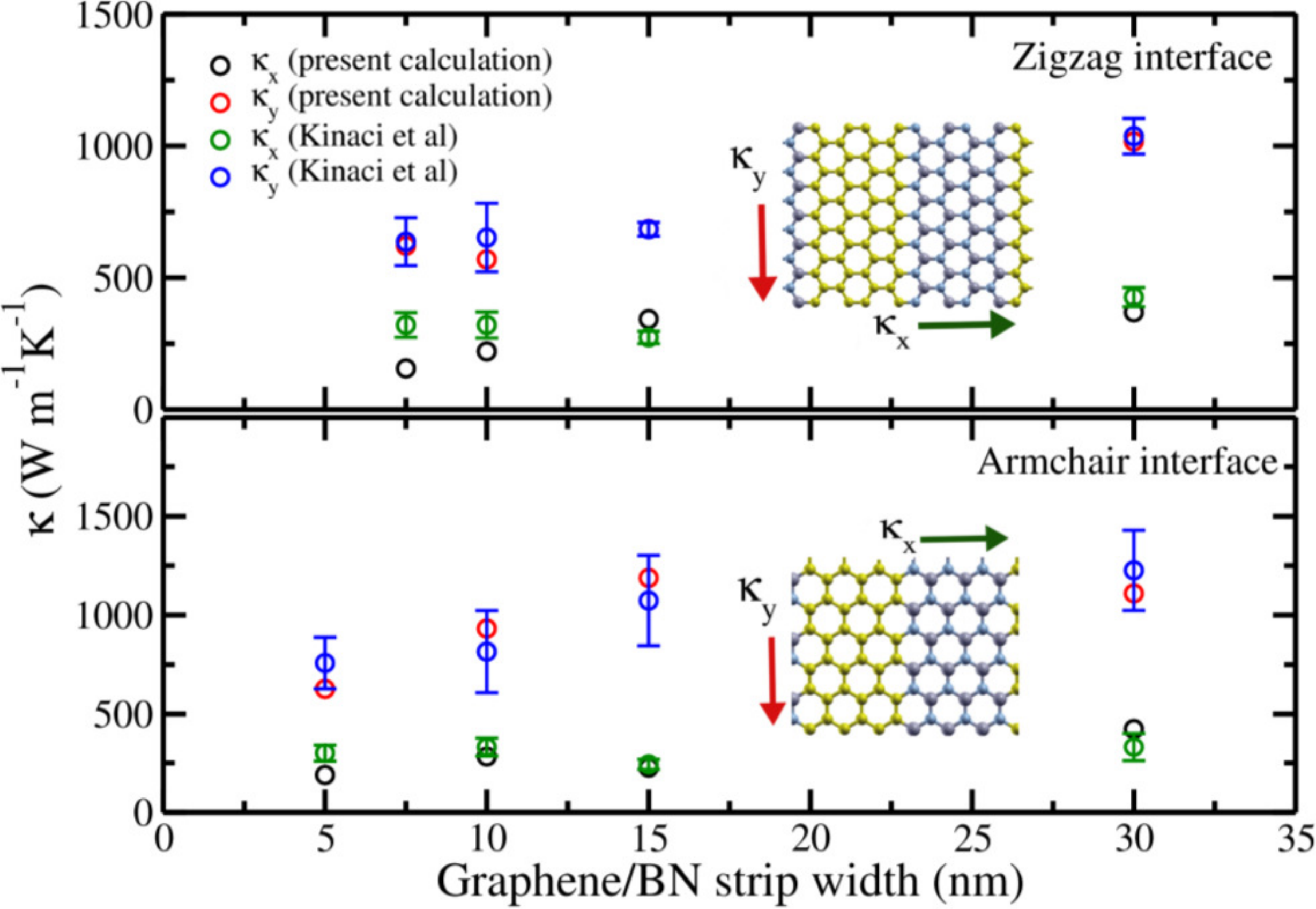}
\centering
\caption{\label{kxky} Calculated parallel and perpendicular components of thermal conductivity of {\it h}-BN/Graphene planar striped heterostructures with zigzag and armchair
interfaces between G and {\it h}-BN domains plotted against the width of the domains. For comparison calculations by Kinaci et al. \cite{kinaci} are also shown. The inset
shows the atomic arrangements in each interfaces.}
\end{figure}

In order to test the equilibrium MD method \cite{lammps} for the calculation of thermal conductivity, we first calculated $\kappa$ for two-dimensional striped 
heterostructures of Graphene and {\it h}-BN with both
armchair and zigzag interfaces between Graphene and {\it h}-BN domains at $T=300$K, shown in Fig. \ref{kxky}. The error bars in our calculation are estimated from five 
different sets of MD runs with different random initial velocities. For comparison we also show the previous calculations by Kinaci et al \cite{kinaci} 
in the same figure. Our simulation results for $\kappa$, both parallel and perpendicular to the crystal edges as shown in the inset, compare quite well with 
previous calculations \cite{kinaci}.
 
We have also compared the ratio $\kappa$/$\kappa_0$, $\kappa_0$ being thermal conductivity of pristine Graphene, and obtained this
to be 0.3203 for the zigzag and 0.3273 for the armchair interfaces, respectively. These results are in excellent agreement with calculations performed using non-equilibrium
Green's function method for Graphene and {\it h}-BN nanoribbons \cite{rubio2012}. Therefore, the use of Tersoff potential based force field was found to be satisfactorily 
applicable for the thermal transport simulations in Graphene and {\it h}-BN based heterostructures. The calculated $\kappa$ for G/{\it h}-BN/G heterostructures for 
different sample thickness are given in Fig \ref{ztT} at different $T$.

\subsubsection{Bulk and multilayers of {\it h}-BN}

\begin{figure}[!htbp]
\includegraphics[scale=0.3]{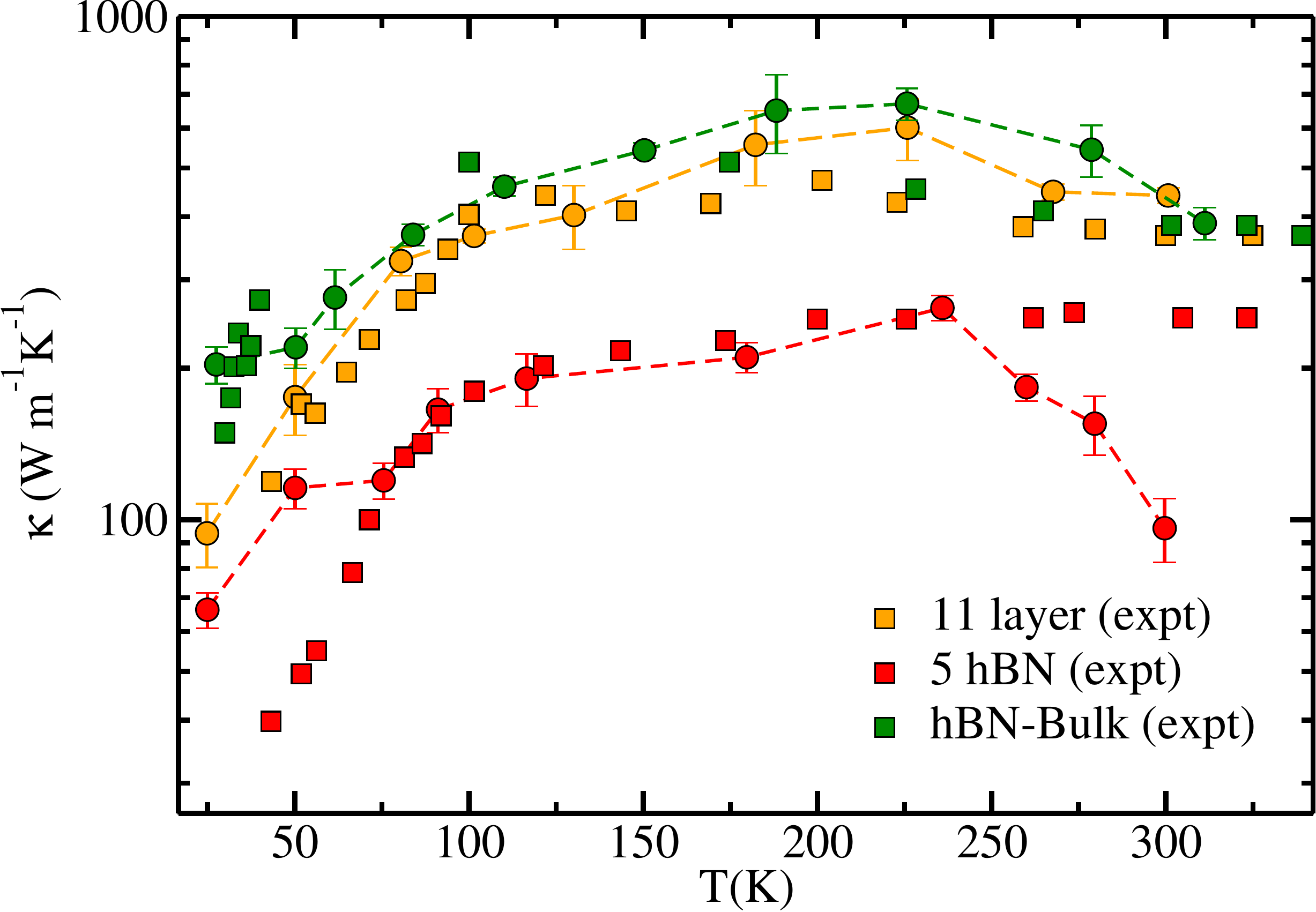}
\centering
\caption{\label{T-Kappa} Calculated temperature dependence of the thermal conductivity of {\it h}-BN layers of different thickness. Experimental results \cite{jo13} are
shown as squares and present calculations as circles. The error-bars are calculated from five different sets of calculations using different seeds.}
\end{figure}

In Fig \ref{T-Kappa} we show the results of thermal conductivity as a function of the temperature (25-400K) calculated from the equilibrium MD simulation at constant
temperature ($NVT$ thermostat)for pure {\it h}-BN 5-, 11-layers and bulk, and compared with the available experimental results of Jo et al \cite{jo13}.
For each calculation, the system was thermalized to the desired temperature first for each set of initial uniform distribution of velocities. $\kappa$ was calculated
for five different sets of initial velocities and the error bar was estimated from the standard deviation.

We found $\kappa$ increases with $T$ and tend to saturate at $T\sim 220$K for each samples of {\it h}-BN, with results for 11-layer tending towards the bulk value. 
Moreover, for each of the three samples
of {\it h}-BN multilayered films, $\kappa$ shows maxima in the temperature region of 200-250K. We observed an overall agreement of our MD simulation results with recent
experimental measurements \cite{jo13}. For 11-layer and bulk {\it h}-BN samples the agreement between the simulation and experimental data is better in temperature
range 25-300K, whereas for the 5-layer sample the MD results seem to agree only in the temperature range 100-250K. 

Recently, several calculations have been reported on thermal conductivity of single-layer {\it h}-BN \cite{lindsay11,mortazavi15,ouyang10}. Transport calculations by 
Lindsay et al \cite{lindsay11} and Ouyang et al \cite{ouyang10}, which were calculated from the phonon spectrum using the phonon Boltzmann transport equation, indicate 
that $\kappa$ shows a maxima around $T\sim 150$K and decreases with $T$. Mortazavi et al  using MD simulations have reported $\kappa$ monotonically decreasing with 
temperature, however all reported values are for temperatures greater than 200K.  Our calculated $\kappa$ for single-layer {\it h}-BN shows a monotonic decrease with 
$T$, not shown here. The numerical value of $\kappa$ and its variation with $T$ depends on the direct and Umklapp phonon-phonon scattering mechanism \cite{lindsay11} 
and also on the lifetime of such processes. We plan to investigate these effects using phonon Boltzmann transport theory from the phonon bandstructure later.

\subsection{Thermal Conductance, Power Factor and Figure-of-merit of G/{\it h}-BN/G Heterostructures}

\begin{figure}[!htbp]
\includegraphics[scale=0.30]{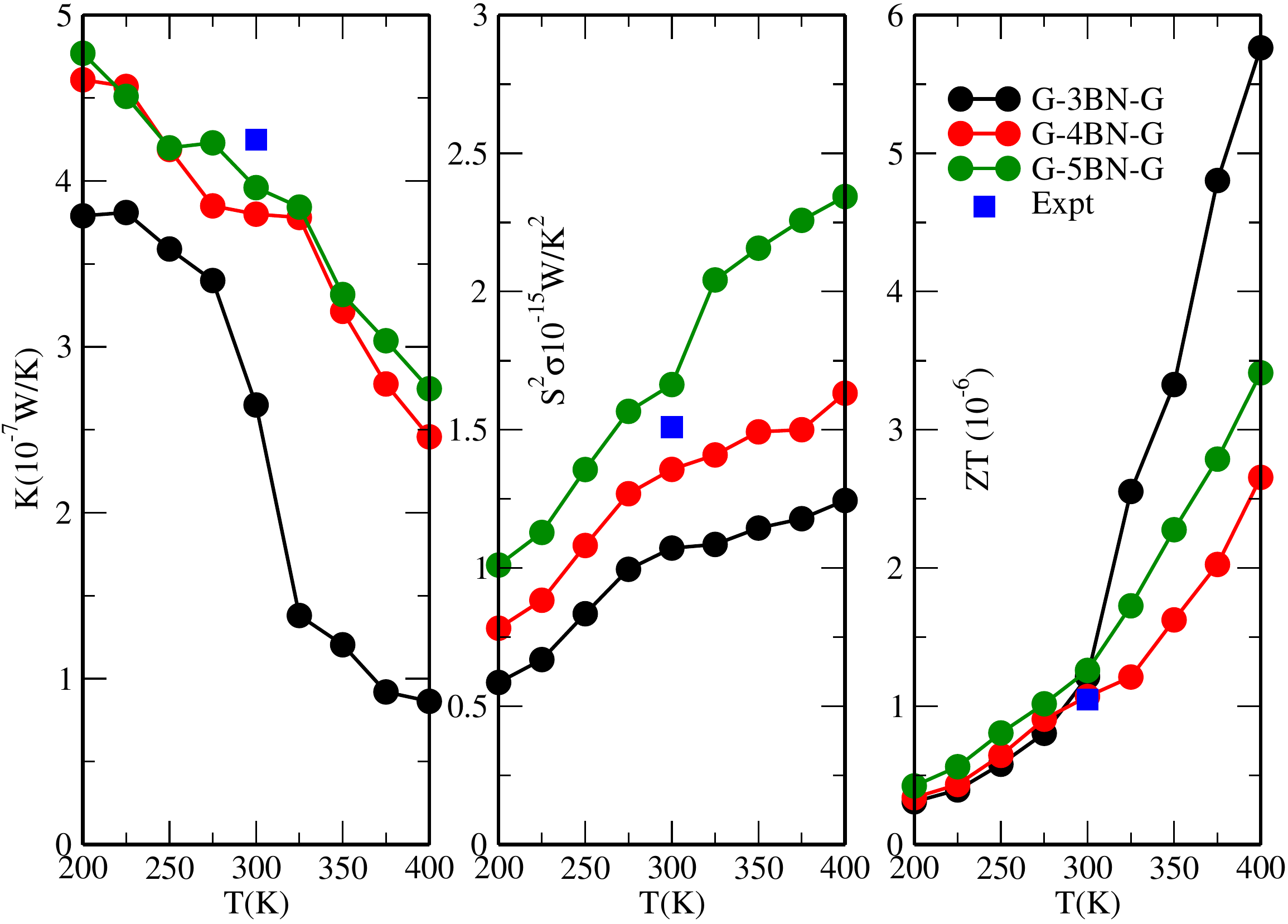}
\centering
\caption{\label{ztT} Calculated temperature dependence of Thermal Conductance ($K$), Power Factor ($S^2G$) and the Figure-of-merit ($ZT$) of {\it h}-BN
layers of different thickness shown in the left, middle and right panels; respectively. The available experimental data \cite{chung14} in each panel at 300K is 
also indicated.}
\end{figure}

The calculated thermal conductance ($K$), Power Factor ($S^2G)$ and the Figure-of-merit ($ZT$) of G/{\it h}-BN/G Heterostructures with three-, four- and five-layers 
of {\it h}-BN at the fixed chemical potential
has been shown in Fig \ref{ztT}. The chemical potential was fixed to obtain the experimentally observed Seebeck coefficient of -99.3 $\mu$V/K as shown in Fig \ref{Smott-fig}. 
Note, we have plotted the thermal conductance in Fig \ref{ztT}, obtained by multiplying the thermal conductivity ($\kappa$) with the height
of the G/{\it h}-BN/G heterostructure as indicated in Fig 1. Similarly, for the Power Factor ($S^2G$), we have taken $G$ as the electrical conductance, so that $ZT$ becomes a
dimensionless quantity. 
It can be seen from Fig \ref {pf} that as we increase the number of layers, the power factor increases whereas the thermal conductivity 
decreases with temperature. As a result the power factor increases with temperature. For G/{\it h}-BN/G heterostructures having 4- and 5-layers, our calculated values 
agree well with the experimental results as the temperature tends towards room temperature.
From equation \ref{ZT}, the Power factor and Figure of Merit will have the same characteristics when plotted against energy at a given temperature, since $\kappa$
is a function of only temperature. 

We would like to emphasize that our calculations involve electrical transport not strictly along the vertical direction, because the Boltzmann transport theory
yields smaller contributions to electrical conductivity along that direction compared to those along $x$- and $y$-directions. We have calculated the $z$-component of the
Seebeck coefficient ($S_z$) \cite{supp} and found this to be finite and comparable to $S_x$ and $S_y$ close to the Fermi energy. However, this could be due to the periodic boundary we have implemented also along $z$- directionwith a vacuum of 22${\AA}$ between the sandwiched layers.
The total Seebeck coefficient $S$ shown
in Fig \ref{Smott-fig}, however, shows a quantitative agreement with the experimental data \cite{chung14}. Thus, we conclude that in the thermoelectric 
measurements \cite{chung14} the electrical transport may not be strictly along the $z$-direction as the upper and lower Graphene contacts with multilayer {\it h}-BN 
would allow transport channels involving components along $x$- and $y$-directions as well. A good quantitative agreement with experimental data also supports above conclusion. 
  
\section{Summary}
We have shown that for three, four and five BN layers sandwiched between Graphene layers, the Boltzmann transport theory gives accurate results for the 
power factor and the Figure-of-merit, comparable to the experimental data. We have also shown that for sufficiently large number of atoms, MD simulations using 
the Tersoff type potential yields results in good agreement with experiments for thermal conductance of multilayer {\it h}-BN, laterally grown striped Graphene and 
{\it h}-BN  two-dimensional heterostructures and sandwiched films of Graphene and multilayered {\it h}-BN, using the equilibrium Green-Kubo method. Our calculations
may be extended to include phonon bandstructure based transport calculations and using non-equilibrium Green's function based methods.

\section{Acknowledgment}
All computations were performed at the High Performance Parrallel Computer platform at S.N. Bose Centre. One of us R.D. acknowledges support from S.N. Bose Centre
through a Senior Research Fellowship.
We would like to thank Prof S.D. Mahanti for insightful discussions and valuable comments.

\end{document}